\documentclass[aps,prl,onecolumn]{revtex4}
\usepackage{graphicx,epsfig}
\usepackage{textcomp}
\usepackage{gensymb}
\usepackage{amsmath}

\begin{document}

\title{Spiral, Core-defect and Wave break in a modified Oregonator Model}

\author
{Parvej Khan and Sumana Dutta {\footnote{sumana@iitg.ac.in}}}

\affiliation{Department of Chemistry, Indian Institute of Technology Guwahati,\\
 Guwahati 781039, INDIA. }





\begin{abstract}
Target waves and spiral waves were discovered in the Belousov-Zhabotinsky (BZ) reaction around 50 years ago. Many biological systems demonstrate such rotating spiral patterns. Spiral waves are widely encountered in the glycolytic activity of yeast and cardiac and neuronal tissues. In the cardiac system, the spiral waves and their three dimensional counterparts, the scroll waves are the cause of arrhythmia. These waves are also widely studied in the experimental BZ reaction as a table-top model of the cardiac system. To understand the underlying physics or chemistry of the system, the Oregonator model is considered in correlation to the BZ reaction system. In this study, starting from the same FKN mechanism, we slightly modify the original Oregonator model. We present a three-variable model that can be called a modified Oregonator model in which we observe the presence of the spiral, spiral defect, and spiral breakup by tuning the parameters.
\end{abstract}



\maketitle

\noindent {P}atterns like spirals and scrolls are abundant in nature. A large number of systems show the formation of rotating spiral waves \cite{win}. From cardiac tissue to BZ reaction-in a wide range of biological, physiological, physical, and chemical systems can be noticed. One of the main ambitions or goals is to understand the electrical activity of the human cardiac system through the study of a spiral and its three-dimensional counterpart \cite{jalife}.

In most of the observed systems, the spiral rotates outwardly from its core, which is known as a pacemaker \cite{copt}\cite{ame}\cite{glyco}. In connection to the cardiac system, a pair of spiral waves in a BZ system are investigated in many ways. The interaction of multiple counter-rotating spiral pairs are being studied by \emph{Kalita et al.} both experimentally as well as in numerical simulation \cite{hk_int}. Spiral and scroll dynamics are being tried to control by different kinds of external gradients like thermal gradients, electric field gradients etc. \cite{luther},\cite{mix}. Spiral wave properties are also being investigated with variation of excitability factor \cite{Dhriti_spiral}.

One of the very interesting as well as important phenomena is spiral break up or wave break up. Spiral breakup is being studied extensively in a computer simulation using CGLE and FHN models \cite{turb_bu}, though we have limited examples in experiments. Sustainable spiral waves in the human heart are comparable with cardiac arrhythmia and the wave breakup is related to more dangerous situations like ventricular fibrillation \cite{bu_vfibri}. So the dynamics of the spiral wave and mechanism of wave breakup is very necessary to study both in experiments also in theory.

Belousov-Zhabotinsky (BZ) reaction is among the most studied oscillatory chemical reactions. It can generate sustainable chemical waves like spirals and scrolls \cite{winfree}. Though there are differences in heterogeneity, conduction velocity, etc., these waves are excitable as the electrical signal of our cardiac system. Theoretical treatment of the mechanistically complex BZ reaction is done with the FKN mechanism and a well-established Oregonator model is often used to generate spiral and scroll waves to study.

In this manuscript, we describe a newly developed model derived from the FKN mechanism, which may also be called as modified Oregonator model. We investigated the model through numerical simulations. We report the formation of spiral, annihilation of spirals and also the exciting wave break phenomena upon parameter tuning. Previously, almost in all studies, specific models are generally used to describe some specific phenomena but here we show almost all kinds of spiral dynamics with this single four-variable reaction-diffusion model.

\section{Model}

\subsection{Mechanism for BZ reaction}

The important steps in the mechanism of the BZ reaction proposed by Field-K\"or\"os-Noyes(FKN) \cite{spiral_scott}, used for development of most numerical models, is given as$-$ 

\begin{eqnarray}
\text{BrO}_3^- + \text{Br}^-+2\text{H}^+ &\xrightarrow{k_1}& \text{HBrO}_2+\text{HOBr} \\
\text{HBrO}_2 + \text{Br}^- + \text{H}^+ &\xrightarrow{k_2}& 2\text{HOBr}  \\
\text{BrO}_{3}^{-} + \text{HBrO}_2 + \text{H}^{+} + 2\text{M}^{2+} &\xrightarrow{k_3}& 2\text{HBrO}_2 + 2\text{M}^{3+}+\text{H}_2\text{O} \\
2\text{HBrO}_2 &\xrightarrow{k_4}& \text{BrO}_{3}^{-}+ \text{HOBr} + \text{H}^{+} \\
2\text{M}^{3+} + \text{MA} + \text{BrMA} &\xrightarrow{k_5}& f\text{Br}^{-} + 2\text{M}^{2+} + \text{Other products}.
\end{eqnarray}

The FKN mechanism involves three main processes: the autocatalysis of HBrO$_2$ (step 3); inhibition by bromide ion (steps 1 and 2),  and resetting the cycle (4 and 5). The values of the rate constants are given as $k_1$= 2 M$^{-3}$s$^{-1}$, $k_2$= 1.25 $\times$ 10$^5$ M$^{-2}$s$^{-1}$, $k_3$= 45 M$^{-2}$s$^{-1}$, $k_4$= 1.4 $\times$ 10$^3$  M$^{-1}$s$^{-1}$, $k_5$= 6.5 $\times$ 10$^{-3}$  M$^{-1}$s$^{-1}$ \cite{spiral_scott}. 

Expressing the reactions in a more general form gives-
\begin{eqnarray}
\textbf{A} + \textbf{W} + 2\textbf{H} &\xrightarrow{k_1}& \textbf{U}+\textbf{P} \\
\textbf{U} + \textbf{W} + \textbf{H} &\xrightarrow{k_2}& 2\textbf{P}  \\
\textbf{A} + \textbf{U} + \textbf{H} &\xrightarrow{k_3}& 2\textbf{U} + 2\textbf{V} \\
2\textbf{U} &\xrightarrow{k_4}& \textbf{A}+ \textbf{P} + \textbf{H} \\
\textbf{V} + \textbf{B} &\xrightarrow{k_5}&{\frac{1}{2}} f\textbf{W}
\end{eqnarray}
This system of equations constitute the Modified Oregonator model, where comparison with the FKN would yield \textbf{A} = BrO$_3^{-}$, \textbf{W}= Br$^-$, \textbf{H}= H$^+$, \textbf{U}= HBrO$_2$, \textbf{P} = HOBr,  \textbf{V} =  oxidized form of catalyst, and \textbf{B} = MA+BrMA (malonic acid and its derivatives). This scheme is able to give, by simple linear combination of equations, the overall reaction,
$f\textbf{A} + 4\textbf{B} + (3f+1)\textbf{H} \rightarrow (3f+1) \textbf{P}$, with no net production or destruction of  \textbf{U}, \textbf{V} and \textbf{W}. Hence, they are considered as the intermediates, whose dynamics is of interest to us.

\textbf{A} and \textbf{B} are major reactants, and their concentrations are treated as constants, \textit{a} and \textit{b}. As the reaction is carried out in an acidic environment, we have an excess of hydrogen ions at any time, and \textbf{H} is also considered as a constant parameter. 


The kinetic behavior of the intermediates can now be written from the above model as-

\begin{eqnarray}
\frac{dU}{d\tau}&=&k_1aWH^2 -k_2UWH +k_3bUH -2k_4U^2\ \\
\frac{dV}{d\tau}&=&2k_3aUH-k_5bV\ \\
\frac{dW}{d\tau}&=&-k_1aWH^2 -k_2UWH + \frac{1}{2}k_5fbV\ 
\end{eqnarray}
Here $U$, $V$, $W$, and $H$ are the concentrations of the respective chemical species in units of mol L$^{-1}$ and $\tau$ is time in units of s.

\subsection{Dimensionless form}
We further non-dimensionalize the equations in the above sequence (11-13), using the following scaling. 
\begin{eqnarray}
u= \frac{k_3}{k_5}\frac{a}{b}U , \hspace{0.5cm}  v=\frac{k_1 k_3}{k_2 k_5}\frac{a^2}{b}V ,   \hspace{0.5cm} 
w=\frac{k_2}{k_3}\frac{1}{a}W ,   \hspace{0.5cm} 
t= k_5 b \tau, \nonumber
\end{eqnarray}
where $u$, $v$, and $w$, are the dimensionless variables, giving us the set of non-dimensional kinetic equations 
\begin{eqnarray}
\epsilon \frac{du}{dt}&=&\frac{1}{2}wh^2-uwh+uh-qu^2\ \\
\frac{dv}{dt}&=& uh-v \\
\epsilon^\mathrm{'} \frac{dw}{dt}&=&-qwh^2-2quwh+2qfv\
\end{eqnarray}
\begin{equation}
\noindent \text{where } h= \frac{2 k_1 k_3}{k_2 k_5} \frac{a^2}{b}H = \frac{1}{h_0} H, \text{is the non-dimensional parametrized form of the } \nonumber
\end{equation}
hydrogen ion concentration, \textit{H}. In the above set of equations, the parameters are given by,
\begin{eqnarray}
\epsilon=\frac{1}{h_0}\frac{k_5}{k_3}\frac{b}{a},  \hspace{0.5cm} 
\epsilon^\mathrm{'}=\frac{1}{h_0^2}\frac{2k_4k_5}{k_2k_3}\frac{b}{a},  \hspace{0.5cm}  \text{and  }\hspace{0.3 cm} 
q=\frac{2k_1k_4}{k_2k_3}. \ \nonumber 
\end{eqnarray}
The values of $a$ and $b$ are considered to meet usual experimental conditions \cite{oreg}, and can be varied to give different values of the parameters. 

Now we have a modified version of the three-variable Oregonator model as equations (14-16).

	.

	
	\section{Simplified two variable model}
	The BZ-system is traditionally studied by the Oregonator model which is derived from the FKN (Field- K\"or\"os -Noyes) mechanism \cite{FKN1972} of the reaction. The Oregonator model is a three-variable system, that is often simplified to a two-variable activator inhibitor model, where bromous acid (HBrO$_2$) is the activator, $u$, and the oxidized form of the catalyst,(Fe$^{3+}$) is the inhibitor, $v$ \cite{oreg}. 
	
	\begin{eqnarray}
	\frac{du}{dt}&=& \frac{1}{\varepsilon}\left(u(1-u)-\frac{fv(u-q)}{(q+u)}\right)\\
	\frac{dv}{dt}&=& u-v
	\end{eqnarray}
	However, the hydrogen ion concentration does not appear in either the two- or three- variable models of the Oregonator, as it gets embedded in the kinetic parameters $f$, $q$ and $\epsilon$. Our modified model described above considers hydrogen ion as an important parameter. The three variable model described above can be more simplified.
	
	As, $\epsilon^\mathrm{'}$ $<<$ $\epsilon$, steady state approximation can be employed on $w$ \cite{oreg}, giving us
	\begin{equation}
\nonumber
w=\frac{2fv}{h(h+2u)}
\end{equation}
	Substituting this value of $w$ in the above equations [14-16], we have:
	
	\begin{eqnarray}
	\epsilon \frac{du}{dt}&=&fv\frac{h-2u}{h+2u}+u(h-qu) \\
	\frac{dv}{dt}&=& uh-v \
	\end{eqnarray}
	
	In the presence of diffusion, the modified Oregonator model takes the following form.
	\begin{eqnarray}
	\epsilon \frac{du}{dt}&=&fv\frac{h-2u}{h+2u}+u(h-qu) +D_u\nabla^2u\\ 
	\frac{dv}{dt}&=& uh-v +D_v\nabla^2v
	\end{eqnarray}
	where $D_u$ and $D_v$ are the diffusion coefficients of $u$ and $v$ respectively.
	
\begin{figure}[h]
	\centering
	\includegraphics[width=.8\linewidth]{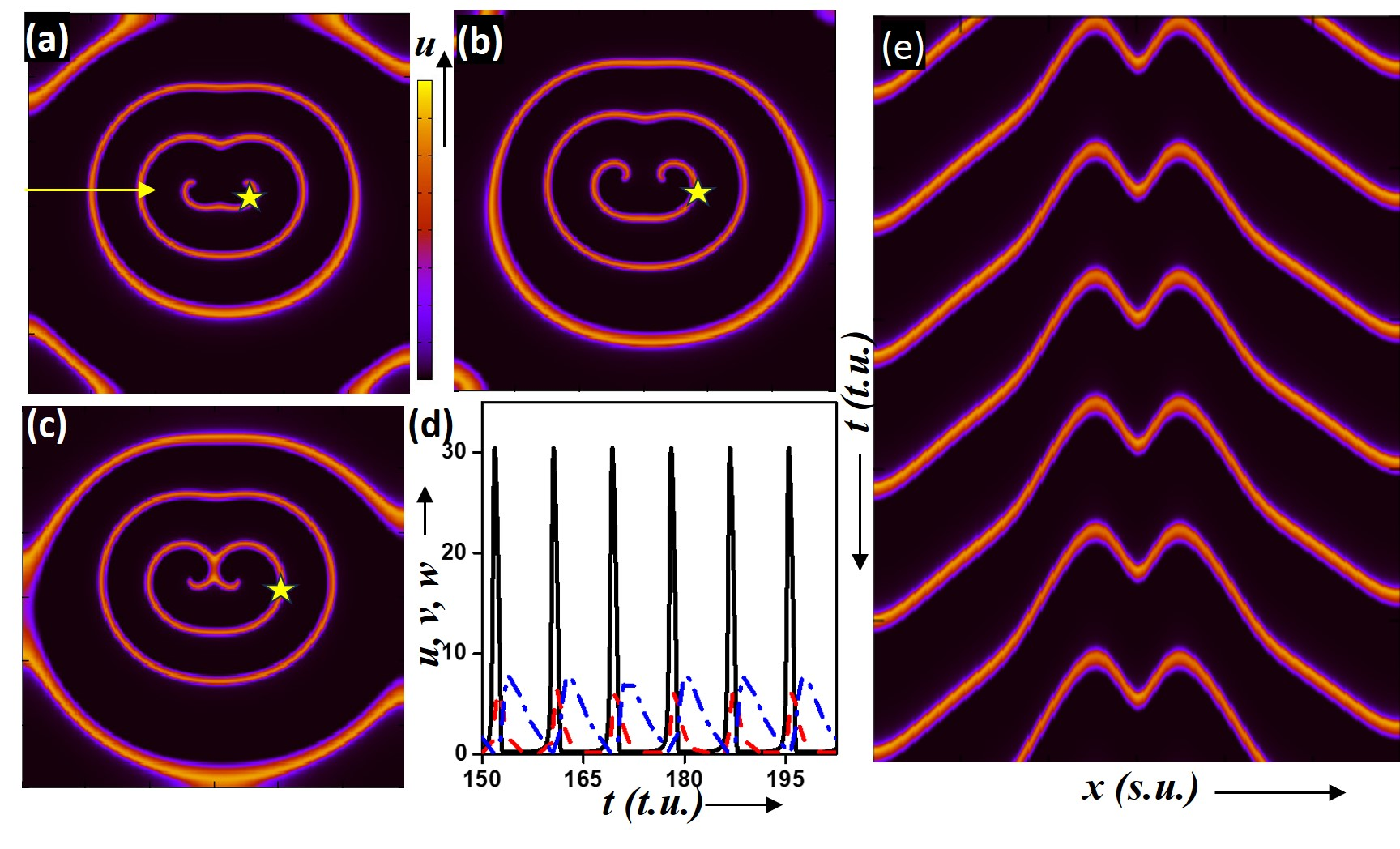}
	\caption{Stable spiral. (a)-(c) Snapshots at 324.2 t.u.,327.1 t.u., and 330.3 t.u., respectively. The star marks on the spiral wave-arm show the outwardly movement of the wave. (d) Time variation of the three variables $u$ as solid black curve, $v$ the dashed red curve and $w$ is the dash dotted blue curve (same curve types are used across the images) (e) Time space plot between 150 to 200 t.u. along yellow arrow shown in (a). Color key for the snapshots and timespace plots is depicted between (a) and (b), and is similar across the figures.}
	\label{fig1}
\end{figure}

\section{Generation of spiral waves}

In order to study spiral wave activity in a spatially extended system, the BZ reaction is often carried out in an unstirred gel medium where diffusion plays a major role in modifying the dynamics of the system. The gel however, rules out any convection in the medium. For such a system, our three variable reaction-diffusion equations are given as:
\begin{eqnarray}
\frac{du}{dt}&=&\frac{1}{\epsilon}\left[\frac{1}{2}wh^2-uwh+uh-qu^2\right] +D_u\nabla^2u\ \\
\frac{dv}{dt}&=&\left[uh-v\right] +D_v\nabla^2v\ \\
\frac{dw}{dt}&=&\frac{1}{\epsilon^\mathrm{'}}\left[-qwh^2-2quwh+2qfv\right] +D_w\nabla^2w\
\end{eqnarray}
where,  $D_u$, $D_v$, $D_w$ are the diffusion coefficients of $u$, $v$ and $w$, respectively.

In order to numerically integrate the equations (23--25), zero flux boundary conditions are applied across all four boundaries. The system parameters are chosen as $\epsilon=0.035$, $\epsilon^\mathrm{'}=0.008$, $f=0.9$, $q=0.01$. The time-step and space-step were chosen as $\Delta t= 0.015$ t.u. and $\Delta x=0.45$ s.u., respectively, which translates to 0.814 ms, and 4.06 $\mu$m. These values of $\Delta t$ and $\Delta x$ are convergent and yield best resolution of patterns in our parameter range, and also can be varied by around 20$\%$, without causing any major changes to the nature of the waves, or the time period and wavelength of the corresponding spirals. Simulations were carried out in a $300\times300$ (135 s.u. $\times$ 135 s.u.) grid. Euler's method is used for the integration, and a central difference method helped to calculate the diffusion term. Stable rotating spirals, with non-meandering circular cores are observed in a range of parameters. For our analysis, we vary the parameters from a base condition, at which the system supports an outwardly rotating spiral. The diffusion constants of $D_u=0.3$, $D_v=0.17$, $D_w=0.34$, with $h=$ 0.3. The values of the diffusion coefficients are taken in a ratio so as to match their actual expected values depending on their molecular weights (molecular weight of $u$ or HBrO$_2$ is 113 g mol$^{-1}$, of $v$ or ferroin is 596 g mol$^{-1}$ and that of $w$ or bromide is 80 g mol$^{-1}$.). The initial values of $u$, $v$ and $w$ are taken to be zero across the entire space, as they are intermediates of the BZ reaction, except for a thin area signifying the initial wave-front. A plane wave is initiated in the middle of the square area, by taking three stripes of width $2 \Delta x$ and length $60 \Delta x$ having non-zero values of the variables, emulating the front and back of the wave. The straight plane wave initiated, curled up to form a pair of counter-rotating spirals \ref{fig1}.

\subsection{Effect of hydrogen ion concentration}
Earlier studies on the BZ system and the Oregonator model had shown that changing hydrogen ion concentration had a major effect on the excitability of the system, and hence modified the wave nature \cite{Dhriti_spiral}. However, in those studies, the effect of the H$^{+}-$ion concentration could only be considered implicitly as a modification of the system parameter, $\epsilon$, or an added acidity factor. Our modified Oregonator model enables us to explicitly vary the initial hydrogen ion concentration to predict the effect of the same on the nature of the system.  

In order to explore the effect of $h$, we start with the base condition, and vary the value of $h$. Increasing the value of $h$ upto 0.55, stable spirals are obtained, beyond which the system has no solution. Decreasing the value of $h$ however yields interesting results. 

\subsubsection{Drifting Spirals}
As $h$ is decreased below 0.29, the spiral starts to get unstable. Now, the spiral tips become anisotropic, and starts drifting away from their initial positions. 

\subsubsection{Spiral breakup}
At still lower $h$ values, the entire system starts getting unstable and wave-break starts initiating far from the core, finally leading to the formation of multiple spirals in the system. The current scenario is reminiscent of the far-field breakup that has been observed in some earlier studies \cite{markusbar}.

\subsubsection{Target pattern}
At $h\le$ 0.17, there is no more formation of spiral. The straight plane wave that we initiate does curl up at its ends, but before they turn further to form a pair of counter-rotating spirals, the two ends touch and form target waves. 

\subsubsection{Chaotic Oscillations and Oscillation Death}
Before we reach oscillation death, at values of $h<0.1$, it passes through a fleeting region of chaotic oscillations, that is distinct from the spiral breakup regions.

\begin{figure}
	\centering
	\includegraphics[width=0.7\linewidth]{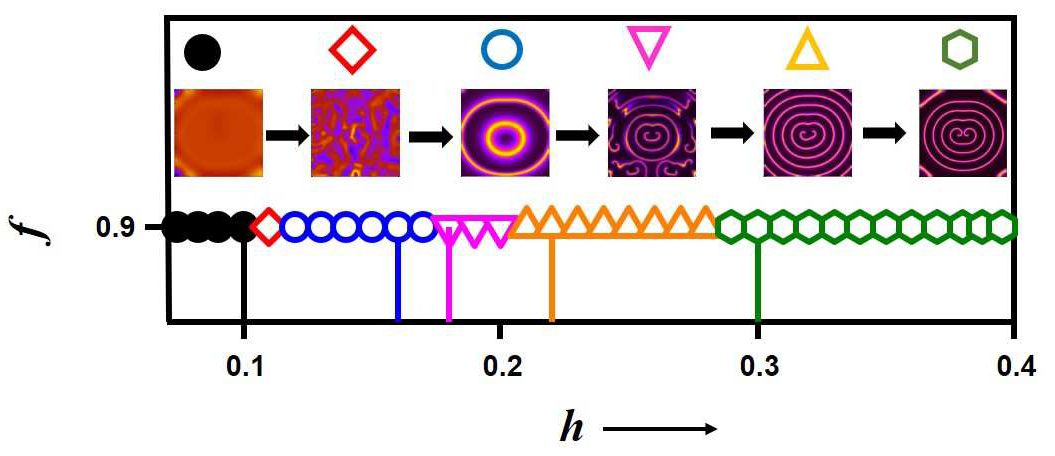}
	\caption{Variation in wave property with $h$ for a fixed value of $f$. Black full circles depict oscillation death, while blue open circles designate Target pattern. Red diamonds stand for chaotic oscillations and magenta upturned triangles denote spiral break-up. The orange triangles are for drifting spirals while olive hexagons are the regular stable spiral waves. The violet squares are for the turbulent dynamics emanating from core defect.}
	\label{fig2}
\end{figure}

For the base condition, a variation in $h$, generates a 1-D phase diagram [\ref{fig2}], showing all the above phenomenon. This suggests that merely changing the hydrogen ion concentration can generate a plethora of phenomena.

Our two variable model described above (Eqs. 21, 22) are also capable of generating spiral waves numerically. The presence of $h$ parameter helps to study the effect of hydrogen ions on the dynamics and wave properties of spirals in a BZ reaction-diffusion system. 

\section{Discussion and Conclusion}
We report the formation of spiral waves, core defect, annihilation of spiral pairs, and the breakup of spirals. Spiral waves have both chemical and biological significance. These are abundant in cardiac tissues, catalytic surfaces (CO oxidation on Pt surface), and aggregation of amoebae. Spiral breakup near the core and outside the core is explained in the paper of Bar \textit{et. al}. They took a more general Barkley model considering a very fast diffusion of the activator and delay in the production of inhibitor. We
obtained both the spiral pair break-up scenario, where – (1) break up starts at the core and then spreads over space, and (2) break up starts away from the core leaving the core greatly unaffected. Many other papers concerned with
the breakup dynamics of spirals also considered the FHN model or, the more general Barkley model with a faster diffusion of the activator \cite{11,13,14,15}.
This current work is different in the sense that it uses a new model and gives a parametric regime of different kinds of spiral dynamics as well as the breakup phenomena. We also considered the diffusion coefficients of inhibitors like Br$^-$ and Fe$^{3+}$ are instead of neglect them like it has been done in other spiral breakup studies with Barkley or FHN
model.

\end{document}